\documentstyle[aps,prb,epsf]{revtex}
\begin{document}

\draft
\preprint{Submitted to {\em Notes and Discussions} in the American Journal of Phyiscs}
\title{Averages of static electric and magnetic fields over a 
spherical region: a derivation based on the mean-value theorem}

\author{Ben Yu-Kuang Hu}

\address{Department of Physics,
250 Buchtel Commons, University of Akron, Akron, OH~44325-4001.
}

\date{\today}
\maketitle
\newpage

The electromagnetic theory of dielectric and magnetic media 
deals with macroscopic electric and magnetic fields, because the microscopic details 
{\bf of} these fields are usually experimentally irrelevant (with certain notable 
exceptions such as scanning tunneling microscopy).  
The macroscopic fields are the average of the microscopic fields over a 
microscopically large but macroscopically small region.\cite{griffiths,jackson}  
This averaging region is often chosen to be spherical, denoted here as
\begin{equation}
\overline{\bf E} \equiv \frac{3}{4\pi R^3}\int_{\cal V} d{\bf r}'\ {\bf E}({\bf r}'),
\end{equation}
where ${\cal V}$ is a sphere of radius $R$ centered at ${\bf r}$
(with a similar definition for $\overline{\bf B}$).  
$R$ is a distance which is macroscopically small but nonetheless large enough
to enclose many atoms.  
The macroscopic $\overline{\bf E}$ and $\overline{\bf B}$ fields obtained by averaging
over a sphere exhibit properties which prove useful in certain 
arguments and derivations.  These properties are as follows:\cite{grif1,jack1}  
\begin{enumerate}
\item if all the sources of the ${\bf E}$-field are outside the sphere,
then $\overline{\bf E}$ is equal to the electric field at the center of the sphere,
\item  if all the sources of the {\bf E}-field are inside the sphere,
$\overline{\bf E} = -{\bf p}/(4\pi\epsilon_0 R^3)$, where 
${\bf p}$ is the dipole moment of the sources with respect
to the center of the sphere,
\item if all the sources of the ${\bf B}$-field are outside the sphere,
then $\overline{\bf B}$ is equal to the magnetic field at the center of the sphere,
\item if all the sources of the {\bf B}-field are inside the sphere,
$\overline{\bf B} = \mu_0{\bf m}/(2\pi R^3)$, where ${\bf m}$ is the 
magnetic dipole moment of the sources.
\end{enumerate}

These results can be derived in a variety of ways.  For example, 
Griffiths\cite{grif1} derives properties 1 and 2 using a combination of results 
from Coulomb's law and Gauss' law, and properties 3 and 4 by writing down the ${\bf B}$-field
in terms of the vector potential in the Coulomb gauge, and explicitly evaluating
angular integrals.\cite{soln-manual}
The purpose of this note is to describe a relatively simple derivation of all four
results, based on the well-known mean-value theorem theorem (described in most
textbooks on electromagnetic theory):
if a scalar potential $\Phi({\bf r})$ satisfies Laplace's equation in a sphere, 
then the average of $\Phi$ over the surface of the sphere 
is equal to $\Phi$ at the center of the sphere;\cite{grif2,jack2} that is, if 
$\nabla^2\Phi = 0$ in a spherical region of radius $r'$ centered at ${\bf r}$, then 
\begin{equation}
\Phi({\bf r}) = \frac{1}{4\pi}\oint d\Omega\ \Phi\Bigl({\bf r} + \
r'\hat{\bf n}_\Omega\Bigr),
\label{avrg-Phi}
\end{equation} 
where $\Omega$ is the solid
angle relative to ${\bf r}$ and $\hat{\bf n}_\Omega$ is the unit vector pointing in 
the direction of $\Omega$. 

Taking the gradient $\nabla$ of both sides of Eq.\ (\ref{avrg-Phi}) with
respect to {\bf r}, we
immediately obtain  
\begin{equation}
{\bf E}({\bf r}) = -\nabla\Phi({\bf r}) = 
\frac{1}{4\pi}\oint d\Omega \ \left[-\nabla\Phi\Bigl({\bf r} + r'\hat{\bf n}_\Omega
\Bigr)\right]= 
\frac{1}{4\pi}\oint d\Omega \ {\bf E}\Bigl({\bf r} + r'\hat{\bf n}_\Omega
\Bigr);
\label{eq3}
\end{equation}
that is, if $\nabla\cdot{\bf E} = 0$ inside a sphere, the average of ${\bf E}$
over the surface of the sphere is equal to the ${\bf E}$ at the center of sphere.
Eq.\ (\ref{eq3}) is the basis of the derivation of the four properties listed above.
To simplify the notation, henceforth in this note it is assumed that 
${\cal V}$ is a sphere of radius $R$ and is centered at the origin ${\bf 0}$.

\begin{center}
{\bf 1. Electric field with sources outside sphere}
\end{center}

When all the static charge sources are outside ${\cal V}$, 
${\bf E} = -\nabla\Phi$ and $\nabla\cdot{\bf E}=0$ inside ${\cal V}$ and
hence Eq.\ (\ref{eq3}) is valid.
We shall now see that 
property 1 follows quite trivially as a special case of Eq.\ (\ref{eq3}) [and hence of
Eq.\ (\ref{avrg-Phi})]. 
 
The average of ${\bf E}$ over ${\cal V}$ can be written as weighted integral 
of the average of ${\bf E}$ over surfaces of spheres with radius $r' < R$
centered at ${\bf 0}$,  
\begin{eqnarray}
\overline{\bf E} \equiv
\frac{3}{4\pi R^3}\int_{\cal V} d{\bf r}'\ {\bf E}({\bf r}') &=& \frac{3}{R^3}
\int_0^R r'^2 dr'\ \left[\frac{1}{4\pi}\int d\Omega\ {\bf E}(r'
\hat{\bf n}_\Omega)\right].
\label{eq4}
\end{eqnarray}
Using Eq.\ (\ref{eq3}) in Eq.\ (\ref{eq4}) immediately yields property 1,
\begin{equation}
\overline{\bf E} =  \frac{3}{R^3}\int_0^R r'^2 dr'\ {\bf E}({\bf 0}) = 
{\bf E}({\bf 0}).
\end{equation}
\newpage

\begin{center}
{\bf 2. Electric field with sources inside sphere} 
\end{center}

For clarity we first prove property 2 for a single point charge inside the
sphere.  The general result can be inferred from the single point charge
result using the superposition principle, but for completeness  
we generalize the proof for a continuous charge distribution.

Utilizing the vector identity\cite{jackson,vector-identity}
\begin{equation}
\int_{\cal V} \nabla\Phi\ d{\bf r} = \oint_{\cal S} \Phi\ d{\bf a},
\end{equation}
where ${\cal S}$ is the surface of ${\cal V}$, 
the average of the 
electric field over ${\cal V}$ centered at ${\bf 0}$ can be written as\cite{jack1}
\begin{equation}
\overline{\bf E} = -
\frac{3}{4\pi R^3}\int_{\cal V} \nabla \Phi({\bf r})\ d{\bf r} = 
-\frac{3}{4\pi R^3}\int_S \Phi\ d{\bf a}.
\label{potsurf}
\end{equation}
Eq.\ (\ref{potsurf}) implies 
the average $\overline{\bf E}$ is determined completely by the potential on the
surface ${\cal S}$.  

We now use a well-known result from the method of images solution for
the potential of a point charge next to a conducting sphere:
the potential on the surface $S$ for a charge
$q$ at ${\bf r}$ inside the sphere is reproduced exactly by an image charge  
$q' = R q/d$ at ${\bf r}'=(R^2/r)\,\hat{\bf r}$ outside the sphere ($\hat{\bf r}$ 
is a unit vector). 
Eq.\ (\ref{potsurf}) therefore implies that the $\overline{\bf E}$ for a point charge 
$q$ at ${\bf r}$ is exactly equal to that of an image point charge $q'$ at
$(R^2/r)\,\hat{\bf r}$. 
But since the image charge is {\em outside} the sphere, 
we can use property 1 to determine $\overline{\bf E}$, 
\begin{equation}
\overline{\bf E} = {\bf E}_{\rm image}({\bf 0}) = 
-\frac{q'\hat{\bf r}}{4\pi\epsilon_0 |{\bf r}'|^2}
 = -\frac{q{\bf r}}{4\pi\epsilon_0 R^3} = 
-\frac{{\bf p}}{4\pi\epsilon_0 R^3},
\end{equation}
where ${\bf p} = q{\bf r}$ is the dipole moment for a single point charge.

{\em Generalization to continuous charge distributions} -- Assume the charge 
distribution inside the spherical
volume ${\cal V}$ is $\rho({\bf r})$.  
A volume element $dV$ at 
${\bf r}$ inside ${\cal V}$ contains charge $dq = \rho({\bf r})\,dV$.  The image
charge element outside the sphere which 
gives the same average electric field as $dq$ is $dq' = 
\rho'({\bf r}')\,dV' 
= dq\,R/r$ at ${\bf r}' = (R^2/r)\,\hat{\bf r}$.
As in the discrete case, the contribution of $dq$ to the average 
${\bf E}$-field in ${\cal V}$ is equal to the electric field at the origin
due to $dq'$,
\begin{equation}
d\overline{\bf E} = d{\bf E}_{\rm image}({\bf 0}) = 
 -\frac{\rho'({\bf r}')\, dV'}{4\pi\epsilon_0 r'^2}\hat{\bf r}
=  -\frac{{\bf r}\, \rho({\bf r})\;dV}{4\pi\epsilon_0 R^3}.
\end{equation}
The averaged electric field due to all charges in the sphere ${\cal V}$ is therefore
\begin{equation}
\overline{\bf E} = \int_{\cal V} d\overline{\bf E}
= -\frac{1}{4\pi\epsilon_0 R^3}\int_{\cal V} dV\ {\bf r}\,\rho({\bf r})
= -\frac{{\bf p}}{4\pi\epsilon_0 R^3},
\end{equation}
where ${\bf p} = \int_{\cal V} dV\ {\bf r}\,\rho({\bf r})$ is the dipole moment with respect
to the origin of all the charges in ${\cal V}$.

\begin{center}
{\bf 3. Magnetic field with sources outside the sphere}
\end{center}

When magnetic field sources are absent in ${\cal V}$, both $\nabla\cdot{\bf B} = 0$ 
and $\nabla\times{\bf B} = 0$, and hence ${\bf B} = -\nabla\phi_M$ where 
$\nabla^2\phi_M = 0$ inside the sphere.  Therefore, the same derivation
for property 1 holds here; that is, the average of the ${\bf B}$-field
over a sphere is equal to its value at the center of the sphere.

\begin{center}
{\bf 4. Magnetic field for sources inside the sphere} -- 
\end{center}

Using the vector potential description of the magnetic field, ${\bf B} = \nabla\times
{\bf A}$, the average over the sphere ${\cal V}$ can be written as\cite{jack1}
\begin{equation}
\overline{\bf B} = \frac{3}{4\pi R^3}\int_{\cal V} \nabla\times{\bf A}
= -\frac{3}{4\pi R^3}\oint_{\cal S} {\bf A} \times d{\bf a}.
\end{equation}
The second equality in the above equation is a vector identity.\cite{vector-identity}
This equation shows that, as in the case of the ${\bf E}$-field and the
scalar potential, the average ${\bf B}$-field over any volume ${\cal V}$
can be computed from ${\bf A}$ on the surface of ${\cal V}$.

We now consider the
contribution to $\overline{\bf B}$ of current element ${\bf J}({\bf r})\,dV$ inside
${\cal V}$.  We can do this by determining the image current element outside the sphere that 
exactly reproduces the vector potential due to ${\bf J}({\bf r})\,dV$ 
on the surface of the sphere.  We choose the Coulomb gauge
\begin{equation}
A_i({\bf r}) = \frac{\mu_0}{4\pi}\int \frac{J_i({\bf x})}{|{\bf r}-{\bf x}|}
\ d{\bf x},
\end{equation}
where $i$ denotes spatial component.  Since 
$A_i$ is related to $J_i$ in the same way as
the electric potential $\Phi$ is to the charge density $\rho$, 
the method of electrostatic images can also be used here to determine the 
image current element. 

The proof of property 4 proceeds analogously to that of property 2.
For ${\bf J}({\bf r})\,dV$ inside the sphere ${\cal V}$,
${\bf A}$ on the surface ${\cal S}$ is reproduced by an image current
element ${\bf J}'({\bf r}')\, dV'= 
(R/r){\bf J}({\bf r})\, dV$,
where ${\bf r}' = (R^2/r)\hat{\bf r}$.
Since the image current is outside the sphere, we can use property 3.  
Thus, the ${\bf J}({\bf r})\, dV$
contribution to the average ${\bf B}$-field in ${\cal V}$ is 
equal to the ${\bf B}$-field at the center due to ${\bf J}'({\bf r}')\,dV',$\cite{irrel}
\begin{equation}
d\overline{\bf B} = d{\bf B}_{\rm image}({\bf 0}) = -\frac{\mu_0}{4\pi} 
\frac{{\bf J}'({\rm r}')\times\hat{\bf r}}{r'^2}\;dV'
= \frac{\mu_0}{4\pi R^3} {\bf r}\times {\bf J}({\bf r})\; dV.
\end{equation}
The contribution for the entire current distribution in ${\cal V}$ is therefore
\begin{equation}
\overline{\bf B} = \frac{\mu_0}{4\pi R^3}\int_{\cal V} dV\ {\bf r}\times{\bf J}({\bf r})
= \frac{\mu_0\,{\bf m}}{2\pi R^3},
\end{equation}
where ${\bf m} = \frac{1}{2} \int_{\cal V} dV\ {\bf r} \times {\bf J}({\bf r})$
is magnetic moment of a current distribution in ${\cal V}$. 

Finally, note that similar arguments hold for charge distributions which are 
constant along the $z$-direction, since the potentials for these satisfy Laplace's
equation in two dimensions, and the method of images is also applicable for line charges.

\begin{center}
{\bf Acknowledgement}
\end{center}

Useful correspondence with Prof.\ David Griffiths is gratefully acknowledged.


\begin{thebibliography}{10}
\bibitem{irrel} The fact that $\nabla'\cdot{\bf J}'({\bf r}')$ may be non-zero is
irrelevant, because we know from Jefimenko's equations (see
Ref.\ \onlinecite{griffiths}, pg.\ 427--429; Ref.\ \onlinecite{jackson},
pg.\ 246--248) that the ${\bf B}$-field
is dependent only {\bf on} ${\bf J}$ and $\dot{\bf J}$ and not $\dot\rho$.  
Retardation effects in Jefimenko's equations can be ignored by assuming that 
the currents were turned on infinitely long ago.  See also Ref.\ \onlinecite{griffiths},
Problem 7.55, pg.\ 339.
\end{thebibliography}
\end{document}